# Influence of itinerancy on the magnetic properties of the single crystals of $Sr_2YRu_{1-x}Cu_xO_6$ solid solutions

Chi Liang Chen[1*], S. M. Rao[1†], Chung-Li Dong[2], Jeng-Lung Chen [3], Yi-Shen Liu [3], C. L. Chang[3], Ting-Shan Chan[2], Hsiang-Lin Liu[4], Hui-Huang Shieh[5], and Maw Kuen Wu[1]

[1]*Institute of Physics, Academia Sinica, Nankang, Taipei,Taiwan.*
[2]*National Synchrotron Radiation Center (NSRRC) , Hsinchu, Taiwan.*
[3]*Department of Physics, Tamkang University , Tamsui ,Taipei County, Taiwan.*
[4]*Department of Physics, National Taiwan Normal University , Taipei, Taiwan*
[5]*Department of Electrical Engerineering, National Defense University, Taoyuan 335, Taiwan*
E-mail: clchen@phys.sinica.edu.tw; rao@phys.sinica.edu.tw, Fax:886-2-27834187 Tel: 886-2-27880058

**Abstract.** X-ray absorption spectroscopy(XAS) Ru L-edge and O K-edge measurements on single crystals of $Sr_2YRu_{1-x}Cu_xO_6$ ($x$=0.01-0.05) solid solutions show that Cu replaces Ru in the lattice which results in a lattice distortion and itinerant holes that increase with increasing $x$. Powder X-ray diffraction (XRD) measurements confirm the lattice distortion through changes in the lattice parameters. Magnetic measurements reveal an antiferromagnetic transition at 32 K in the crystals with x=0 and a diamagnetic transition at 32 K for x ≥ 0.01indicative of superconductivity in the zero field cooled (ZFC) curves. The field cooled (FC) curves show a weak ferromagnetic character with x=0 and a ferromagnetic transition at 30 K and an antiferromagnetic transition at 23 K in the presence of Cu. The peak at 30 K in the ac susceptibility measurements indicates a spin-glass behavior. The low field part of the M-H curve for x=0.05 crystal indicates weak superconductivity. The shifting of the M-H curve on FC indicates spin-glass state. In conjunction with the XAS results the magnetic behavior is explained in terms of a spontaneous vortex state as a result of the non-freezing $Ru^{5+}$ spins.

## 1. Introduction

A large number of Rutheno-cuprates have been investigated in recent years for co-existing magnetism and superconductivity [1-5]. Among these the family of compounds with the general formula $Sr_2RuLnCu2O8$, Ln=Gd, Y, Dy, Er (generally referred to as Ru-2112) have a Cu-O plane as in the YBCO type compounds that is responsible for the observed superconductivity. On the other hand the family of compounds belonging to $A_2ReRuO_6$, A=Sr, Ba and Re=Y, Ho, Pr (2116) exhibit superconductivity with as little as 0.05 mol Cu replacing Ru. Among the later the $Sr_2YRu_{1-x}Cu_xO_6$ (SrY2116 for brevity) solid solutions discovered during a systematic investigation of the $YSr_2Cu_3O_{6+}$ compound by replacing Cu with Ru [5] have been extensively investigated by magnetic, neutron diffraction and muon-spin rotation studies and found to be superconducting after annealing at 1380-1410 °C in an atmosphere of mixed O:Ar in the ratio 9:1 [4-6]. The parent compound $Sr_2YRuO_6$ is a well known antiferromagnetic (AFM) insulator crystallizing in the $P2_1/n$ space group [7]. To explain the observed superconductivity in this system in the presence of even such a small amount of Cu Ren and Wu [8] proposed a model according to which replacement of Ru by Cu brings about itinerant holes which in turn produce a lattice distortion involving a 'double exchange' mechanism proposed by de Gennes [9]. The model assumes that when $Cu^{3+}$ incorporates in the lattice twice as much $Ru^{6+}$ is formed in the lattice for charge balance. With a resultant spin 1, $Ru^{6+}$ promotes long range superconducting order. The lattice distortion and itinerancy predicted by the Ren-Wu model were indeed reported recently by Chen *et al* [10] from the room temperature x-ray absorption spectroscopy (XAS) measurements on SrY2116 single crystals. Only $Cu^{1+}$, $Cu^{2+}$ and $Ru^{5+}$ were observed in these measurements. Raman spectroscopic measurements have been used to reveal the presence of nano-scale secondary phases in ferroelectrics [11]. Our measurements [12] did not show the presence of either the superconducting $YSr_2Cu_3O_6$ [13] or $Sr_2YRuCu_2O_8$ [14] phases reported in the literature [13]. A systematic investigation of single phase SrY2116 single crystals containing $x$=0-0.05 was therefore made by XAS, magnetic and calorimetric measurements to probe into the influence of Cu on the magnetic properties of SrY2116. Through a correlation of the magnetic and electronic properties it is concluded that while Cu incorporation produces itinerant holes that may pair to produce superconductivity at low temperatures, the non freezing $Ru^{5+}$ spins (3/2) do not permit a long range superconducting order leaving the crystals in a superconducting spin-glass state. This finding in itself may provide an evidence for the peculiar nature of superconductivity in this system as secondary superconducting phases would not be influenced by the Ru spins. While superconductivity in this system has been contested with results on samples that were not subjected to the above treatment [13], a more systematic investigation on well characterized single crystals appears to yield the rich physics that this system offers.

## 2. Experimental

The starting powders were prepared and crystal were grown as per the details described earlier.[15,16] It may be pointed out that extreme care had to be taken regarding the purity of starting materials and even with the solvent $PbO-PbF_2$ as background impurities were found to give rise to negative moments reported in literature [17]. Negative moments were observed even in the case of $x$=0 crystals with 5-10 Oe external field that became positive at higher fields. While doped crystals were grown with >99.9% pure solvents (PbO and $PbF_2$), >99.99% purity powders were used for selected undoped crystals (due to cost considerations). The powder x-ray diffraction (XRD) patterns of the crystals in powder form were recorded using Phillips diffractometer. The magnetic properties were measured using a SQUID Magnetometer (Quantum Design make) and resistivity measurements were made using a physical property measuring system (PPMS) of the same make. The XAS Ru L-edge and O K-edge measurements were carried out with synchrotron radiation at the National Synchrotron Radiation Reach Center (NSRRC) in Taiwan. The beam line 16A (used for Ru *L*-edge spectra) has a resolution of ~0.4 at 2.8keV, and the beam line 20A (used for O K-edge) has a resolution of ~0.1 at 530eV. All the XAS spectra were recorded at room temperature (RT), in the total electron yield (TEY) mode for O K-edge and in fluorescence yield (FY) mode for the L-edge. All spectra were normalized to a unity step height in the absorption coefficient across the edge. The standard metal foils and oxide powders of CuO and $Cu_2O$ were used for energy calibration for Cu and $RuCl_3$ and $RuO_2$ were used in the case of Ru.

## 3. Results and Discussion
### 3.1 XRD measurements

All the crystals used in the present experiments were taken from the same batches of crystal growth to ensure identical composition and conditions of growth. The XRD patterns for SrY2116 crystals shown in Fig. 1 for $x$=0.0, 0.02 and 0.04 have been fitted to the space group $P2_1/n$ and indexed. The peaks are observed to shift to a higher 2θ value as $x$ is increased (31° peak shown expanded in inset (i) for clarity) indicating a decrease in lattice parameters. Additional peaks are not observed in the patterns indicating that secondary phases are not formed during crystal growth. The lattice parameters calculated from these XRD patterns are given in Table 1. While the *a* and *b* parameters tend to decrease with increasing $x$ while the *c* parameter is found to increase. The angle β is also found to increase at the same time. The lattice parameters of the $x$=0 crystals are very close to the reported values [7, 18]. Recent Raman spectroscopic measurements on these crystals also confirm these lattice parameter changes on Cu incorporation through changes in bond lengths [12].

### 3.2 XAS measurements

The Ru $L_3$-edge spectra of the SrY2116 crystals with $x$=0.0-0.05 taken from the same batches of crystals used for XRD measurements are presented in Fig. 2 which show two peaks at 2840 and 2843 eV marked A and B respectively, that correspond to the $t_{2g}$ and $e_g$ bands representing the transitions from $2p_{3/2}$ to $4d$ orbitals of Ru and agree with the published data [10,19-21]. The peak positions are similar to those reported for $Ba_2YRuO_6$ [19]. The spectra were recorded on freshly cleaned crystals to avoid surface contamination effects that might affect the spectra as the penetration of the low energy photon beam is relatively low and have been normalized to the highest peak at 2843 eV. It is seen that the intensity of the $t_{2g}$ peak reduces as $x$ is increased (inset Fig. 2 (i)). This is also confirmed by calculating the area under the peaks A and B. The ratio A/B is found to decrease as $x$ is increased (inset Fig. 2 (ii)). The reduction of A/B ratio means an increase of the $e_g$ band as the spectra are normalized to the peak $C_2$. This in turn indicates decrease in the O-Ru-O bond angle [10]. The present results have been obtained on a series of crystals grown in air with $x$=0.0~0.05 that contain a slightly higher composition of $Cu^{1+}$.

The O K-edge spectra recorded on the SrY2116 crystals exhibit three main peaks as shown in Fig. 3. The one at 528eV corresponds to the Ru $t_{2g}$ ($D_1$) band and the two broad peaks at 531.3 and 532.3eV ($D_2$ and $D_3$ respectively) correspond to the Ru $e_g$ band. These peaks represent the hybridization of the Ru $t_{2g}$ and Ru $e_g$ bands with the O $2p$ orbitals in the undoped crystals [21-25]. A slight decrease in the $t_{2g}$ band height and an increase in the $e_g$ band width is observed with increasing Cu doping. This indicates a decrease in the O-Ru-O bond angle and an increase in the hybridization of the Ru $t_{2g}$ and Ru $e_g$ bands with the O $2p$ bands. It is also seen that the $D_3$ peak height is slightly higher than the $D_2$ peak indicating a higher $Cu^{1+}$ compared to $Cu^{2+}$ as argued by Chen *et al.*[10]. The $D_3$ peak also shifts to a slightly higher energy indicating and increase in the Jahn-Teller (J-T) $\Delta E_{J-T}$ splitting of the $e_g$ band as $x$ is increased. This is an indication of the decrease in the $a$ and $b$ parameters and increase of the $c$ parameter [10, 26] which is in agreement with the XRD data above.

The decrease in the Ru $t_{2g}$ band in the Ru *L*-edge spectra is found to be much smaller than that of the Ru $t_{2g}$ band in the O K-edge spectra. This means that the increase in the unoccupied states is coming from Cu and not Ru. This being the case, the hybridization observed in the O K-edge spectra through an increase in the intensity of the $D_2$ and $D_3$ peak in the $e_g$ band may be coming predominantly from the $Cu^{1+}$ and $Cu^{2+}$ multi scattering. The ground state of the nominal $Cu^{2+}$ may be represented as a hybridization of $3d^9$ and $3d^{10}\underline{L}$ states where $\underline{L}$ denotes a hole on the O $2p$ valence band i.e. a legand hole. Similarly $Cu^{1+}$ would represent a combination of $3d^94s^1$ and $3d^{10}4s^1\underline{L}$ in the ground state [10, 27-29]. These mobile holes are also found to increase with increasing $x$. The influence of these itinerant holes on the magnetic properties will be examined in the next section.

### 3.3 Magnetic measurements

The ZFC and FC magnetization (M) vs temperature (T) curves with 10 Oe external field are presented in Fig. 4(a) and (b) for SrY2116 crystals with $x$=0.0 to 0.05

Fig. 4(b) gives the expanded view of the low temperature region showing the transitions. The figure presents a few interesting features: (1) the ZFC curve for the $x$=0 crystal exhibit two peaks at 32 K and 21K; (2) the ZFC curves for $x$=0.01 crystals shows a diamagnetic transition with a peak at 32 K. As the Cu content increases it is found that this peak decreases and a diamagnetic transition that increase in magnitude as well as temperature with increasing $x$ is observed. (3) while the ZFC curves are indicative of superconductivity the FC curves indicate a ferromagnetic (FM) transition at 32 K and AFM transition at 23 K; (4) the ZFC and FC transitions occur at the same temperature and (5) an upturn is seen in the ZFC curve and a down turn is seen in the FC curve at 23 K in Cu doped crystals.

### 3.3.1 Confirmation of Antiferromagnetic behavior of the crystals

We first examine the behavior of the $x$=0 crystal in more detail. Investigation of the polycrystalline samples of $Sr_2YRuO_6$ sintered at high temperatures by Battle and Macklin [7] revealed an antiferromagnetic (AFM) state with a $T_N$ of 26 K and a spin-flop transition at ~12 K with high external field. In the present case two peaks are observed at slightly higher temperature in the ZFC curve with low external field (10 Oe) presented in Fig. 4(a) for comparison with the Cu doped crystals. ZFC and FC M-T curves obtained on crystals oriented with the *a-b* plane and *c*-axis parallel to 5 kOe external field are shown in Fig 4(c) where the ZFC curve and FC curves show an irreversibility at 25 K. The ZFC curve in the case with the field applied along the *a-b* plane show a peak at 22 K and a shoulder at 29 K while for crystals with same orientation Cao *et al* [17] report a broad and flat peak in this range of temperatures. A single peak is observed at 27 K with the filed along the *c*-axis. On the other hand with several small crystals of x=0 put together two peaks are observed at 22 K and 29 K as is the case with poly crystalline samples [7. 16]. We will not go into the details of these here but to indicate the similarity of the result obtained on polycrystalline samples. These results confirm the phase purity of the crystals used in the present investigations. These results negate the claim of Singh et al [16] on the negative moment observed in undoped bulk SrY2116 in both low and high fields. Their 1kOe field curve is similar to the ZFC curve obtained with x=0.05 crystal shown in Fig 5(a). It is likely that the data represents the influence of impurities in the starting powders used in preparing the samples.

A fitting of the high field data (for the crystal with the a-b plane oriented parallel to external field) (Fig. 4(c)) in the paramagnetic region to the Curie-Weiss law yields a $\mu_{eff}$ of 3.23 $\mu_B$ and θ of -42 K. While the effective magnetic moment is close to the value reported by Cao *et al* [17], the $\theta/T_N \sim 2$ agrees well for a normal antiferromagnetic. It is seen that the 10 Oe curves for *x*=0 (Fig. 3(b)) exhibit irreversibility at 35 K and a large difference between the ZFC and FC curves indicating weak ferromagnetism and a spin-glass state [17, 30, 31].

*3.3.2     Diamagnetic transition in the presence of Cu*
Next we examine the diamagnetic transition observed in crystals with $x \geq 0.01$ which is indicative of superconductivity as reported by Chen *et al* [4, 5]. As *x* is increased the diamagnetic transition increases in magnitude and shifts to a slightly higher temperature. This is indicative of an increase in the superconducting fraction as a result of the increase in the mobile holes as discussed above. To confirm that this is not from the grain boundary conduction or secondary phases the crystal was crushed to a fine powder after recording the M-T curve and the M-T curve of the powder was recorded (given in the supporting data). The magnitude of the signal dropped by 40% only as is expected for powders indicating that this is a bulk feature.

To probe further into this negative moments, M-T curves were recorded on a *x*=0.05 single crystal oriented with the *a-b* plane parallel to different applied fields as shown in Fig. 5(a). It is seen that as the field increases, a peak similar to the one observed in the *x*=0.01 crystal (Fig. 4(b)) appears to develop at 100 Oe field and keeps increasing in magnitude and at the same time shifting to lower temperature as the field is further increased. Fig. 5(b) shows the ZFC and FC curves for *x*=0 and 0.05 crystals oriented with the *c*-axis parallel to the external field of 1 kOe where a single peak is observed at 29 K in the case of the *x*=0.05 crystal and with *x*=0 in addition to this a peak is observed at 20 K. The FC curve with *x*=0.05 shows a ferromagnetic transition at 34 K and a peak at 22 K indicative of an antiferromagnetic ordering in a ferromagnetic background i. e. a canted magnetic state as suggested by de Gennes [9] which is not observed in the *x*=0 crystal. Similar observation reported by Harshman *et. al.* [32] on bulk SrY2116 samples with *x*=0.1 with 3.3 kOe fields was attributed to spin-glass behavior. In the present crystals the diamagnetic signal persists even at 1 kOe indicating that this is not due to remnant field effects. Felner *et. al.* [33] have attributed the appearance of a negative moment in Eu-2122 after a maximum in the positive moment to a spontaneous vortex state, in which vertices caused in the superconducting (SC) plane from the internal filed ($B_{int}=4\pi M$) due to the FM in the Ru lattice. In the SrY2116 crystals also the FM interaction may occurs between the next nearest neighbor Ru ions [7] when the long range AF order is disturbed in the presence of Cu as seen from Fig. 4 above where the AF transition with *x*=0 changes to diamagnetic one with *x*=0.01 in the ZFC curve and FM one in the FC curves. The FM contribution may also arise from the lattice distortion that may result from itinerant holes observed above as per the model of Ren and Wu [8]. Following Felner *et al* [33] we may attribute the negative moment in the $x \geq 0.01$ crystals to SC state. The absence of zero resistance in the crystals led us to look further for possible reason.

To gain more insight into the negative moments some classroom like experiments were performed given in the supporting data. In the first experiment equal weights of undoped SrY2116 crystals and small pieces of as grown (without oxygen annealing) single grain $SmBa_2Cu_3O_{6+\delta}$ were mixed together and used for measuring the M-T curves. In the second experiment small pieces of $SmBa_2Cu_3O_{6+\delta}$ and a commercial ferrite were put together and the M-T curves recorded. The first gave a typical superconducting M-T curve where as the second gave a curve (given in the supporting data) similar to the doped SrY2116. The stepped behavior of the curves may be a result of the domain wall motions in the magnet. The M-H curve in the inset of the figure (is different from the doped SrY2116 crystals) shows a negative slope initially but shows a positive moment at higher fields and the curve is representative neither of a ferromagnetic nor a superconductor character.  These experiments demonstrate that if the SC phase is separated from the antiferromagnetic phase a superconducting signal is seen. Otherwise the M-T curve shows a signal similar to a SC-FM trapped state. The observed behavior of the SrY2116 crystals may therefore be explained in terms of vortex state discussed above involving an interconnected superconductor and ferromagnetic states probably due to small superconducting clusters.

### 3.3.3 Spinglass investigations

These crystals were also examined by some other tests like the AC susceptibility, specific heat (Cp), ZFC and FC M-H curves in addition to the M-T curves with different fields presented above for possible spin-glass behavior [30] as presented below.

The real part M' of the ac susceptibility measured as a function of temperature is shown in Fig. 6. The measurements were made with zero dc magnetic field and 17 Hz ac field at a frequency of 5 kHz in the temperature range 5-300 K. The curve exhibits a peak at 31 K. The inset shows measurements performed close to the peak at three different frequencies 1, 5 and 10 kHz. As the frequency increases the peak magnitude decreases while shifting to a slightly higher temperature. This indicates the presence of spin-glass state as the M' exhibits a peak at the spin-glass freezing temperature as described by Maydosh [30]. But the ZFC dc magnetization curves presented in the inset of Fig. 5(a) do not show either a positive moment or a cusp for x=0.05 crystal even at 50 Oe as observed by Wang *et al* [34]. Therefore, the peak in the present ac susceptibility data may not be due to a simple spin-glass phase and needs more investigations to trace the origin.

The M-H curves of the x=0.05 crystal shown in Fig. 4(a) appears as a near straight line (paramagnetic or AF like) with the origin in the negative side of the moment that becomes positive above 1 kOe. Low field part of the M-H curve of the x=0.05 crystal given in the inset (i) shows an initial negative slope with a minimum ($H_{C1}$) at 6 Oe indicating the presence of a weak superconducting state. An oscillatory behavior in the intial part of the M-H curve may be attributed to the presence of superconducting cluster: spin-glass state as proposed in superconductor composites by Ebner and Shroud [35]. The short period of oscillation is an indication of small size of the cluster as discussed by these authors. The M-H curve recorded after cooling the crystal with a field of 2000 Oe shifts to positive side. Weak ferromagnetism observed in the M-H curves of *x*=0 crystals given in inset (ii) is not seen in the Cu doped crystals. These results also point to the presence of a spin-glass state [29] in the crystals while negative slope in the low field M-H curve showing $H_{c1}$ indicates the presence of a weak superconducting state. The absence of weak ferromagnetism (observed in the *x*=0 crystal) in the *x*=0.05 may result from the vortex state discussed above in a canted magnetic state. Based on this we may conclude that the crystals may represent a superconducting spin-glass as argued by Rao *et al* in the case of $Ba_2PrRu_{1-x}Cu_xO_6$ crystals [29] similar to a spontaneous vortex state.

### 3.3.4 Specific heat measurements

The specific heat ($C_p/T$ vs T) curve presented in Fig. 8 shows a broad hump between 32 and 20 K. However, on applying a field of 6T a drop is observed in the curve at high temperature (~32 K) part of the bump while the lower temperature portion is not affected suggesting the presence of a FM [36] or SC [37] state at high temperature as the Cp peak decreases with filed in both cases and an AF state at low temperature which is not influenced by external field. An upturn is also seen below 5 K in the presence of magnetic field indicating the presence of magnetic clusters. This is consistent with the Ren-Wu theory which predicts at least two peaks of which the higher temperature peak would reduce in the presence of magnetic field [8]. Two sharp peaks were reported by Wu *et. al.* [38] as suggested by this models that were attributed to the SC and AF transitions by comparing with the magnetic data.

The results presented above namely the diamagnetic transition in the ZFC curves, ferromagnetic transition in the FC curves, the AF peak at 27 K in the FC curves and the reduction of Cp at high temperature are in agreement with the propositions of the Ren-Wu theory [8]. However, the theory requires that Cu incorporation as $Cu^{3+}$ resulting in $Ru^{6+}$ in the lattice in an amount that is twice that of $Cu^{3+}$. The XAS measurements [10] show the presence of only $Ru^{5+}$.

To compare the results obtained here with the $RuSr_2GdCu_2O_8$ (2112) system cited above, where magnetism arises from the Ru $t_{2g}$ states of the $RuO_6$ octahedra and do not influence the superconductivity since the Ru $t_{2g}$ electrons do not couple with the Cu $e_g$ states irrespective of the ordering of the Ru moments [39-40]. Since the $RuO_6$ octahedra are rotated by about 14° around the *c*-axis in this system, electronic and magnetic structure are sensitive to crystal structural distortion, but have little influence on the superconducting properties that are caused by the Cu $e_g$ band structure. On the other hand, in case of the SrY2116 system, Cu replaces Ru in the octahedral sites causing a structure distortion in addition to the Jahn-Teller (J-T) distortion due to the perovskite structure. Cu incorporation is also found to increase unoccupied states in the Ru $e_g$ band due to hybridization of Cu and O as presented above. The itinerant holes so produced may pair to give superconductivity as observed in the diamagnetic transition in the magnetic data which is consistent with the model proposed by Ren and Wu [8]. However, as mentioned above the $Ru^{5+}$ with spin (3/2) does not allow the onset of a long range superconducting order. On the other hand some of the $Ru^{5+}$ is converted to $Ru^{6+}$ in the presence of $Cu^{3+}$ (each $Cu^{3+}$ produces twice the number $Ru^{6+}$ [8] in the lattice) when the samples are subjected to annealing at 1400-1420 °C in an atmosphere of Argon and Oxygen. In the presence of $Ru^{6+}$ with spin 1 a larger superconducting volume fraction is produced to give zero resistance. Thus the predominance of $Ru^{5+}$ with spin 3/2 in the as grown crystals does not permit large superconducting volume fraction but probably small isolated clusters that lead to a superconducting spin-glass state [15, 30]. While the long range AF ordering is disturbed in the presence of Cu, the resulting canted magnetic state appears to dominate the magnetic properties in the as grown crystals.

## 4. Conclusion

In conclusion, XAS investigation of the SrY2116 crystals shows that Cu replaces Ru in the lattice causing a lattice distortion and itinerancy that may lead to superconductivity according to the model of Ren and Wu [8]. While the tests like ac susceptibility, field dependent M-T curves and ZFC and FC M-H curves indicate a spin-glass state in the crystal, a diamagnetic transition and negative slope in the low field part of the M-H curves with a $H_{c1}$ at 6 Oe indicate the presence of superconductivity. The Cp decreases at the temperature of the diamagnetic transition as proposed by the model due to the FM state. Based on these results it is concluded that the crystals may represent a spontaneous vortex state due to the canted magnetism as proposed by Ren and Wu [8].

**ACKNOWLEDGMENTS**

This work was supported under National Science Council (NSC) of the Republic of China grant # NSC-93-2112-M-001- 0044. We are very grateful to Prof. H. C. Ren for giving very good insight into the Ren-Wu model and the canted state in this system, Prof. J. K. Srivastava for very useful discussions, M. J. Wang and N. Y. Yen for help in crystal growth, XRD and SEM measurements, and M. C. Ling for Raman measurements. One of the authors (S. M. R.) is grateful to the IOP, Academia Sinica and National Science council of ROC for financial support and C. L. C thanks NSC for support under NSC99-2112-M-001-036-MY3. Ling-Yun Jang, J. M. Chen, and J. F. Lee for the experimental support from NSRRC in Taiwan are gratefully acknowledged.

**Figure Caption**

Fig.1 (a)XRD patterns of SRY2116 as grown $x$=0~0.04; The patterns shift to higher $2\theta$ as $x$ increases indicating a decrease of lattice parameters; inset gives an enlarged view of the main diffraction peak at 31° showing the shift to higher $2\theta$.

Fig. 2 The Ru L-edge spectrum recorded on SrY2116 $x$=0~0.05. Inset (i) gives expanded view of the peak A to show the decreasing of the peak with increasing x. Inset (ii) gives the ratio of the peaks A/B as a function of x and inset (iii) gives the expanded view of the peak B to show no change in intensity as the curves are normalized to this peak.

Fig. 3 The O K-edge spectra of the SrY2116 $x$=0~0.05 crystals. The peak $D_1$ decreases in height and the band width of the $D_2$ and $D_3$ peaks increase as x is increased.

Fig.4 (a) ZFC M vs T curves for $x$=0~0.05 crystals grown in air, inset gives the FC M-T curves (b) an expanded view of the transition region to highlight the transitions; (c) M-T curves of $x$=0 crystals oriented with the *a-b* plane and *c*-axis parallel to 5 kOe field along with several small crystals randomly oriented to the same field.

Fig. 5(a) M-T curves of $x$=0.05 crystal measured with different field from 10~1000 Oe with the *a-b* plane parallel to the field, inset (i) gives the ZFC and FC curve at 1000 Oe and inset (ii) gives the ZFC and FC curves together for a $x$=0.03 crystal with H=10 Oe; (b) M-T curves recorded on $x$=0 and 0.05 crystals oriented with the *c*- axis parallel to 1k Oe external field.

Fig. 6 The real part M' of of the ac susceptibility of a SrY2116 x=0.05 crystal measured as a function of temperature. The inset gives an expanded view of the peak position showing the decrease in M' as frequency is increased as well as a shift to a slightly higher temperature. Dashed line is to guide the eye.

Fig. 7 M-H curves of x=0.05 crystal ZFC and 2000 Oe FC. Inset (i) gives the low field data showing a negative initial slope with a Hc1, Inset (ii) gives the ZFC and 500 Oe FC M-H curves. All curves are recorded at 5K.

Fig. 8 Cp/T vs T curves for a x=0.05 crystal with zero and 6T external fields. A dip is observed in the curve from ~ 32 K where the diamagnetic and FC transitions occur in the M-T curves.

**Table I**

| $x$ | $a$ Å | $b$ Å | $c$ Å | $\beta$ |
|---|---|---|---|---|
| 0 | 5.7829 | 5.7712 | 8.1182 | 89.7 |
| 0.01 | 5.7703 | 5.7700 | 8.1377 | 90.1 |
| 0.02 | 5.7651 | 5.7684 | 8.1529 | 90.16 |
| 0.04 | 5.7626 | 5.7693 | 8.1678 | 90.34 |

It is seen from the table that the lattice parameters of $x=0.0$ crystal are very close to those reported [7,18]. While $a$ and $b$ tend to decrease as $x$ increases, $c$ and $\beta$ are found to increase indicating a lattice distortion in the presence of Cu.

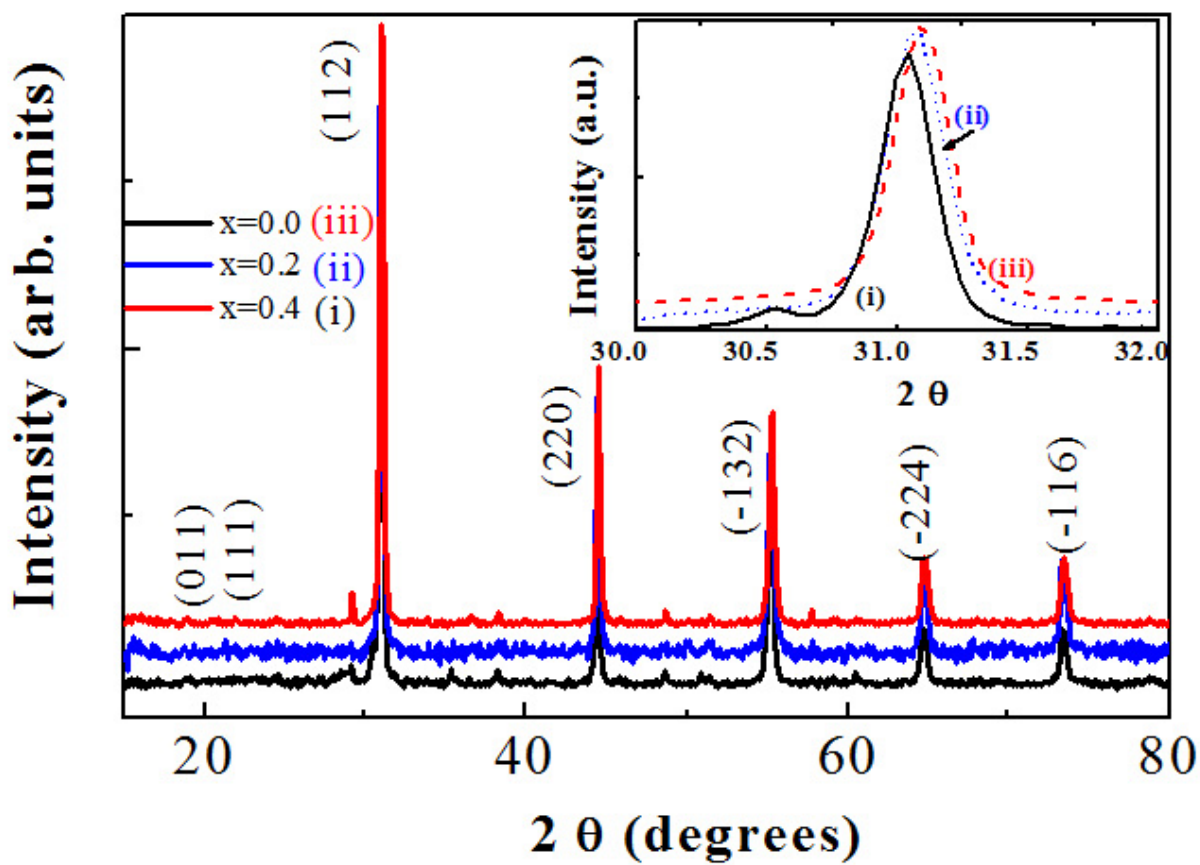

Figure 1

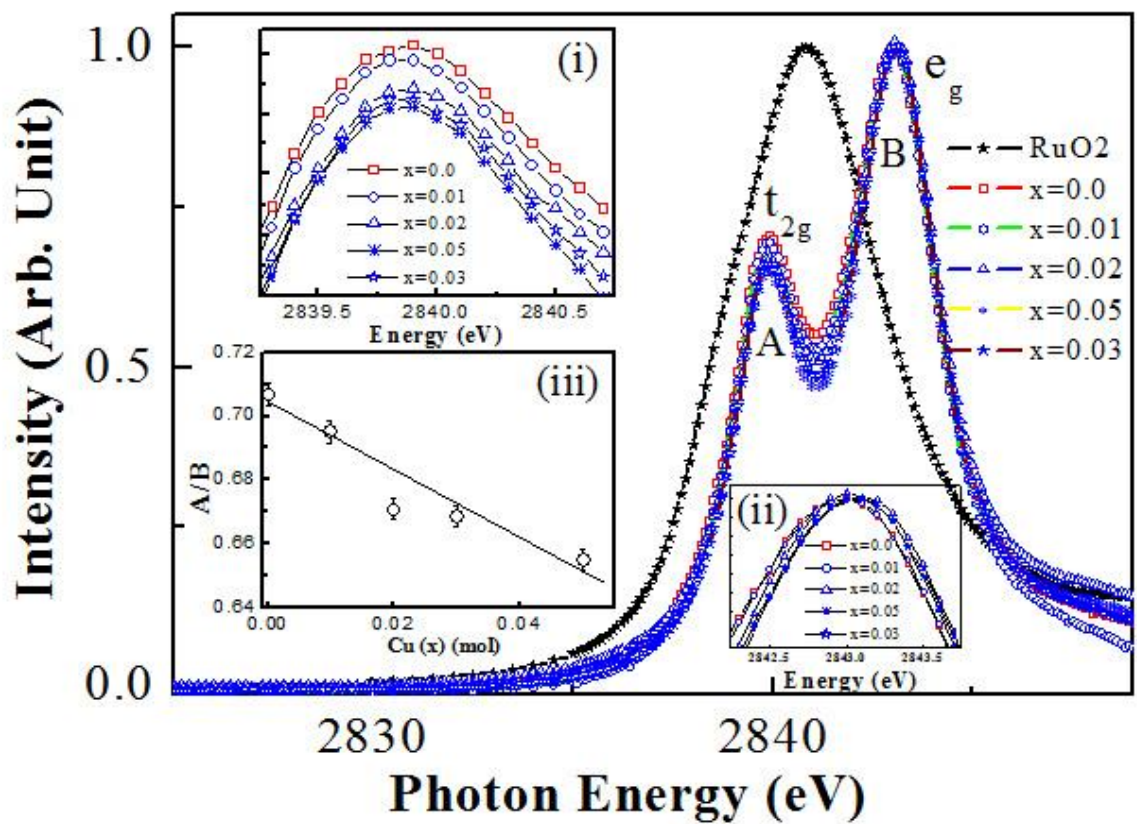

Figure 2

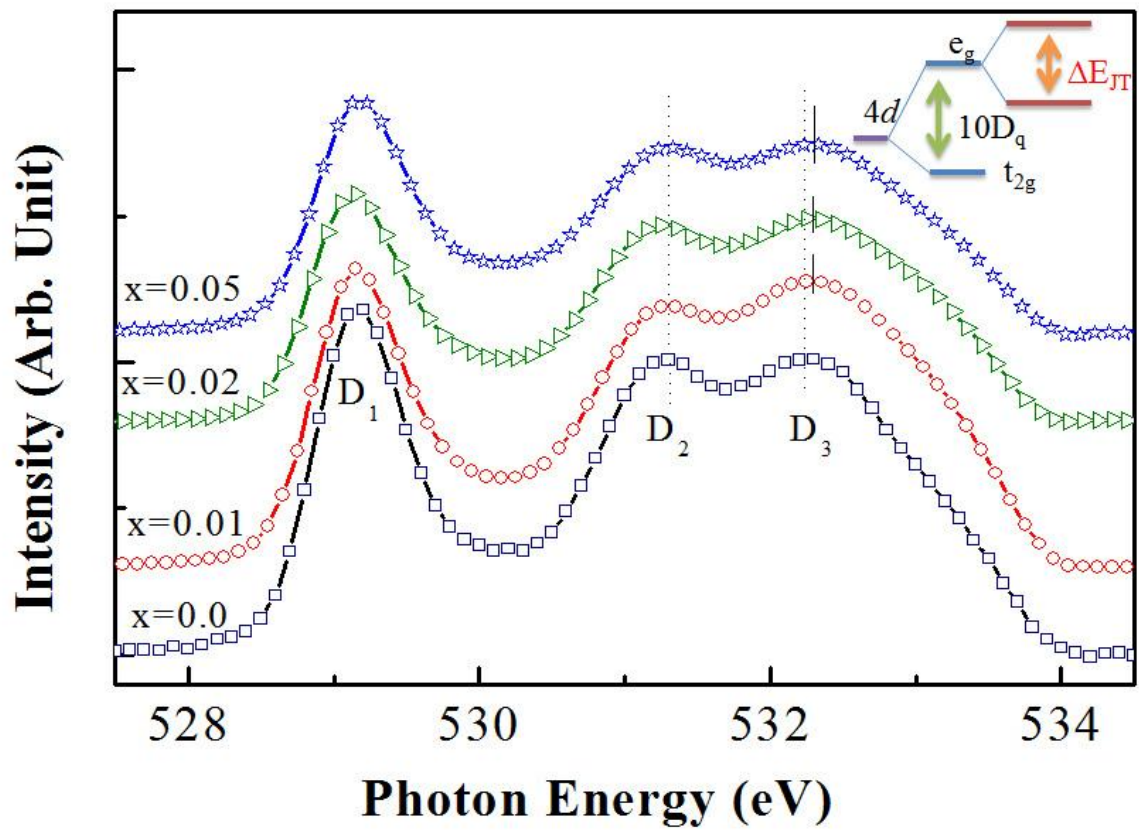

Figure 3

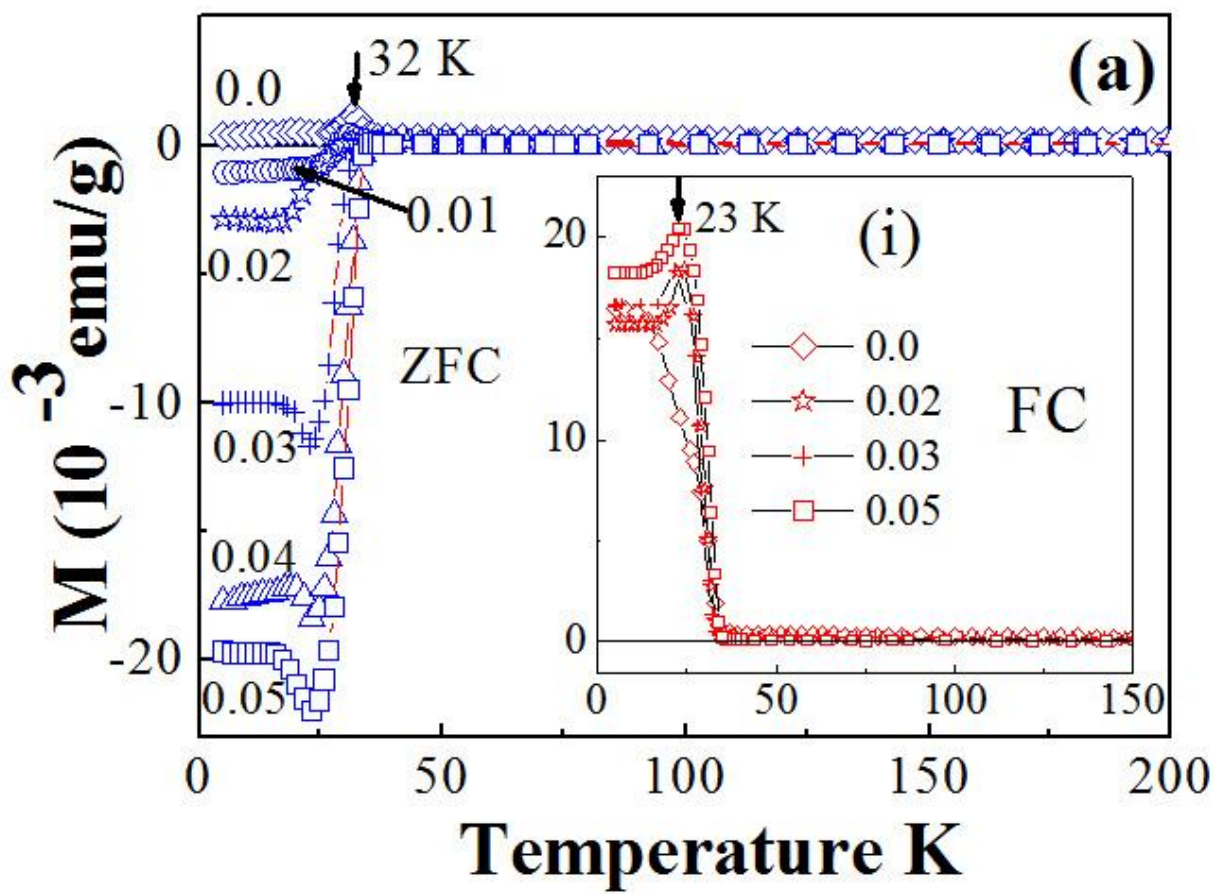

Figure 4(a)

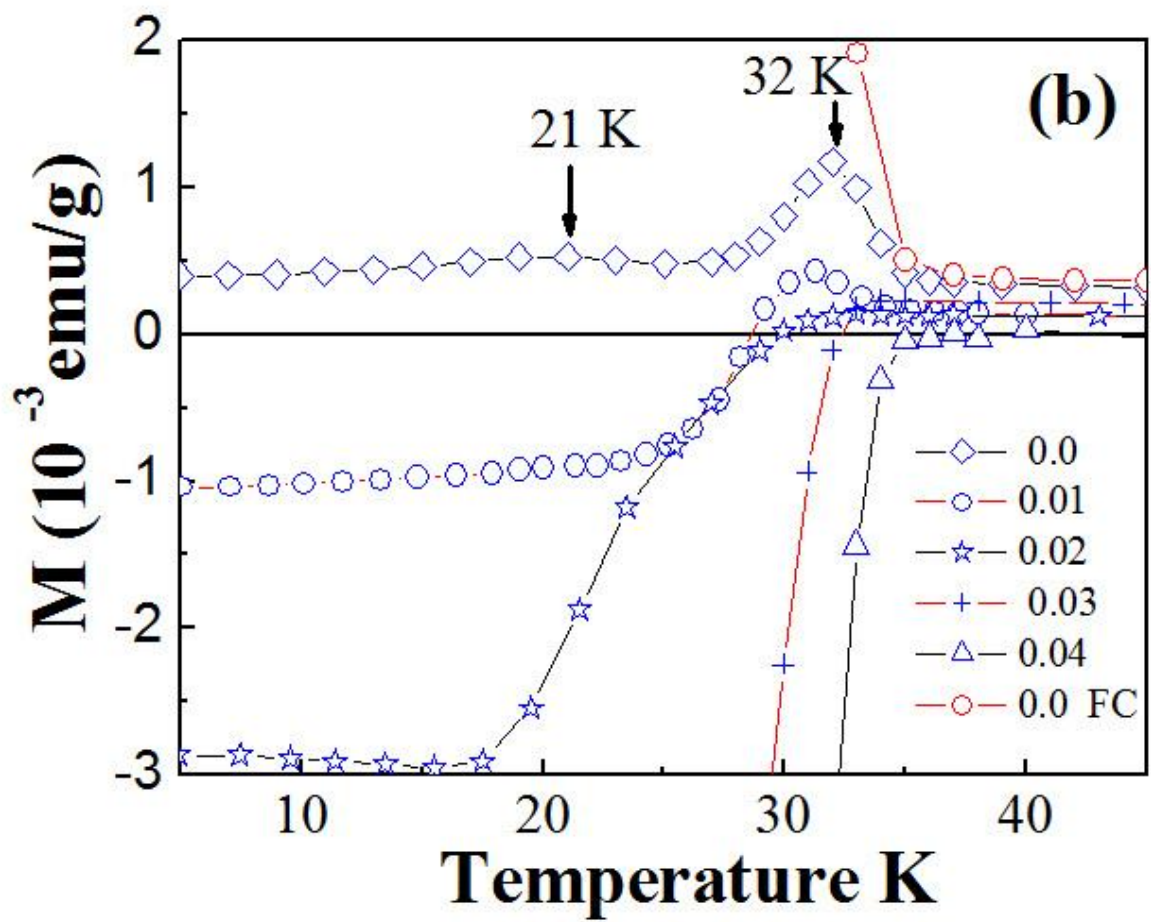

Figure 4(b)

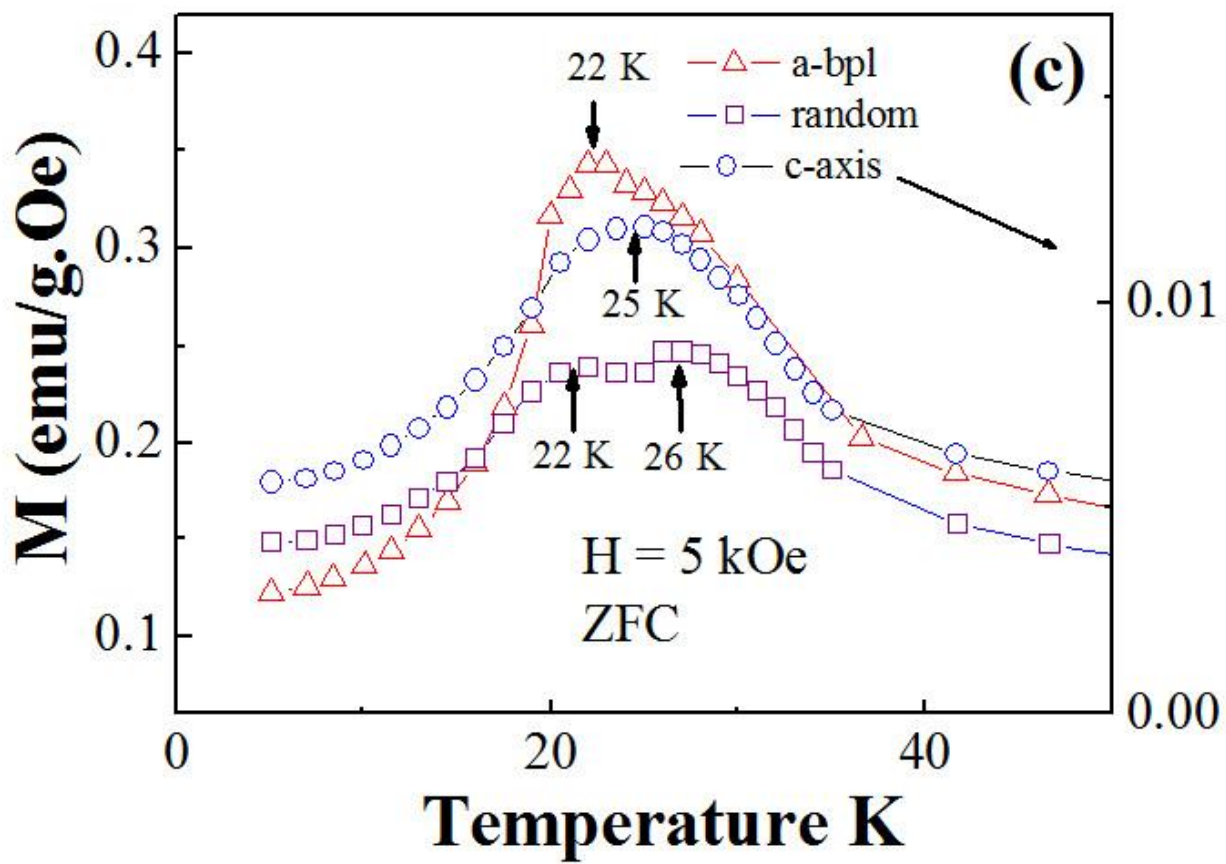

Figure 4(c)

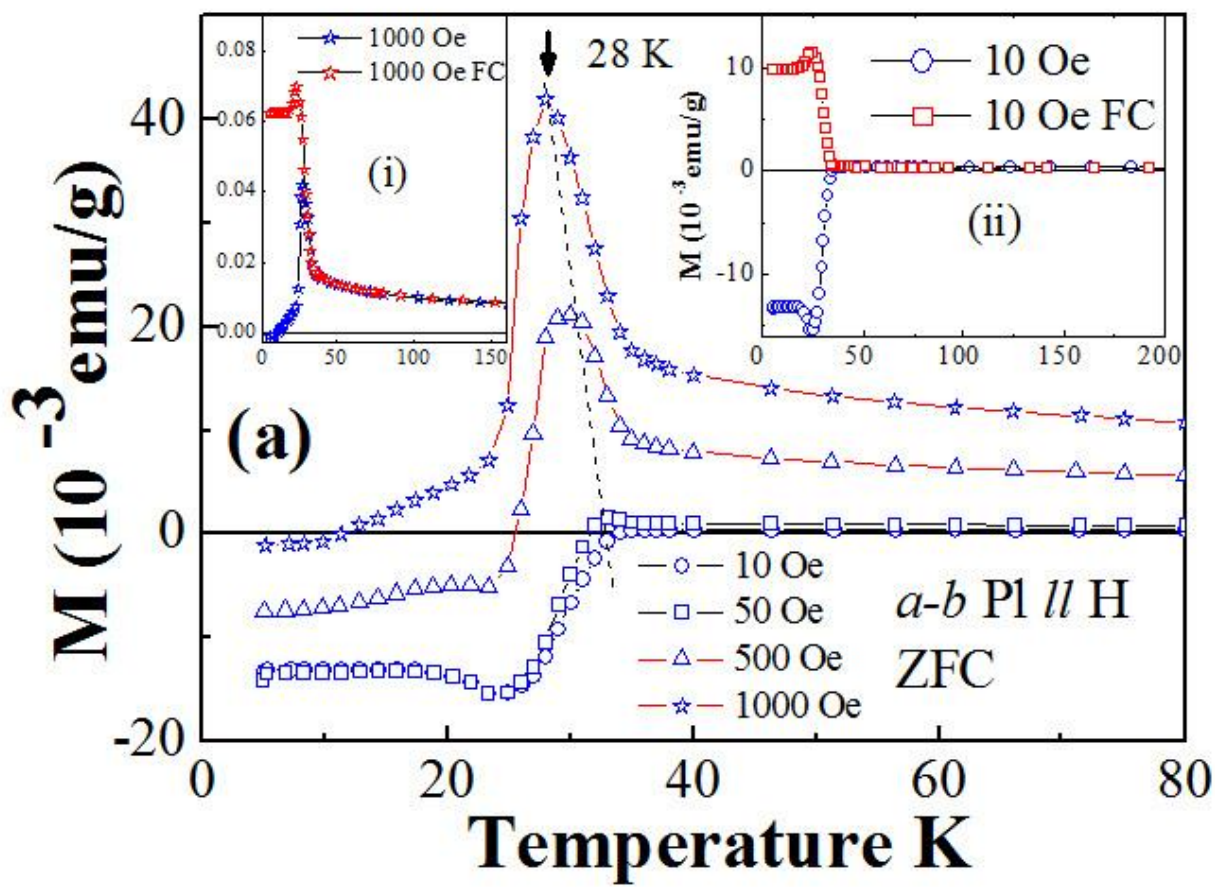

Figure 5(a)

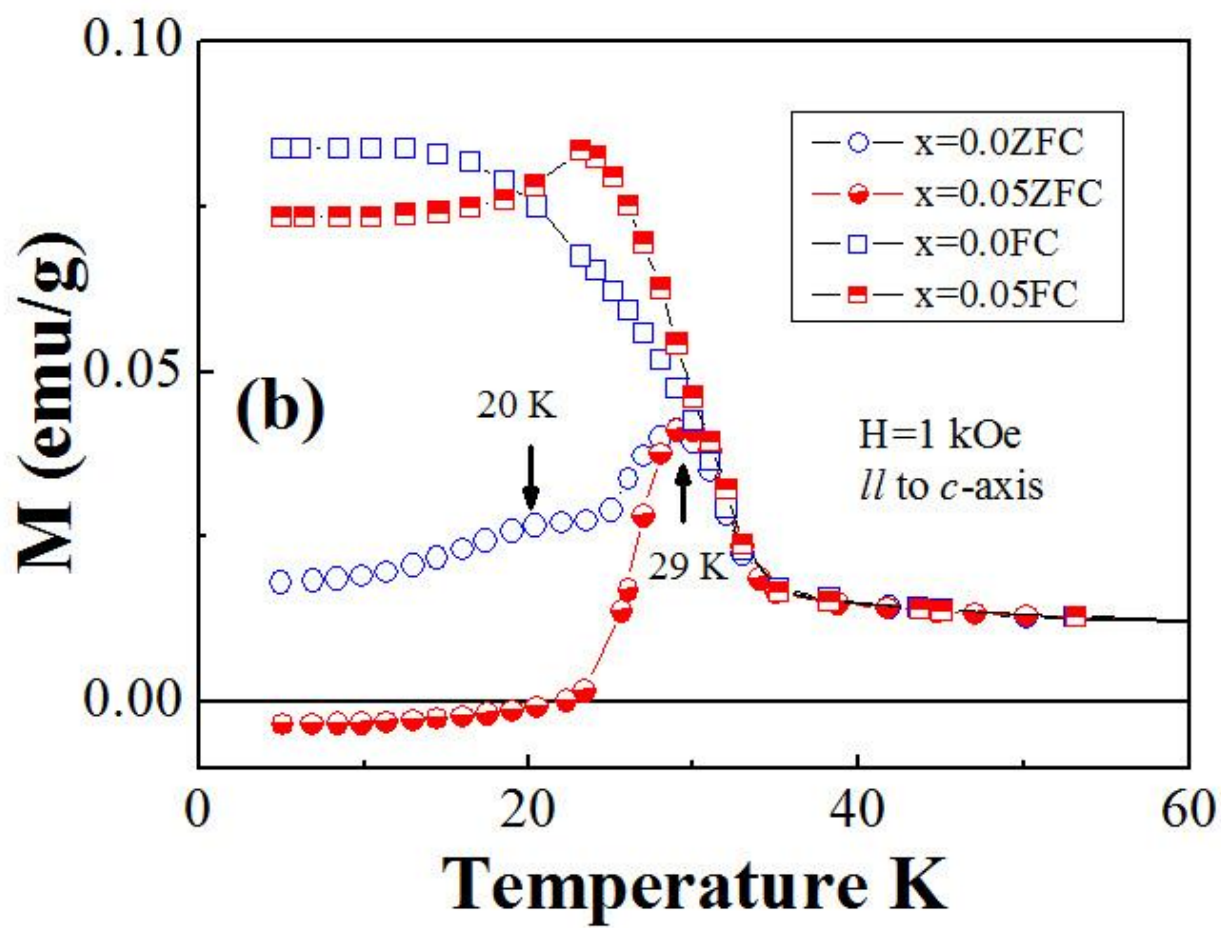

Figure 5(b)

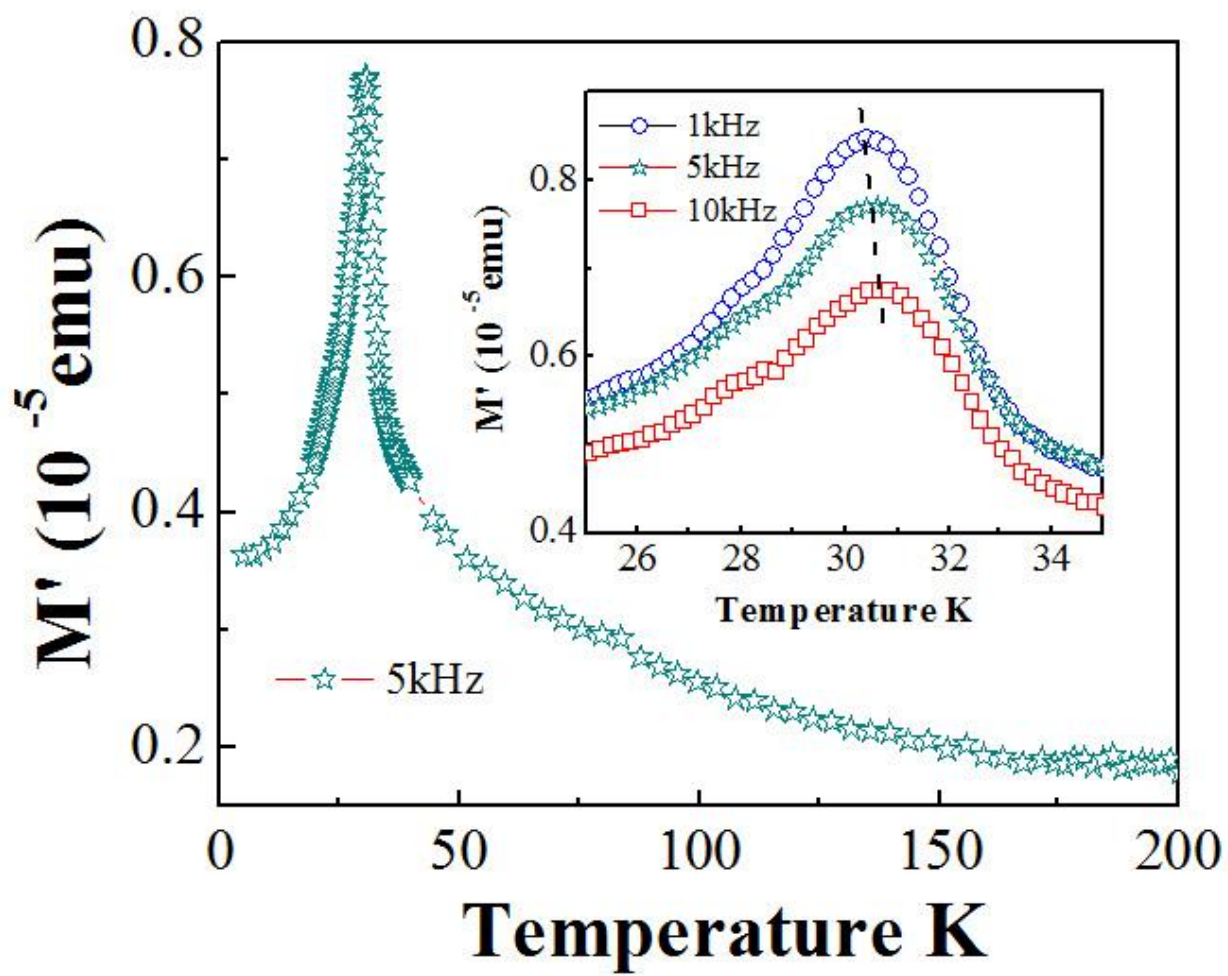

Figure 6

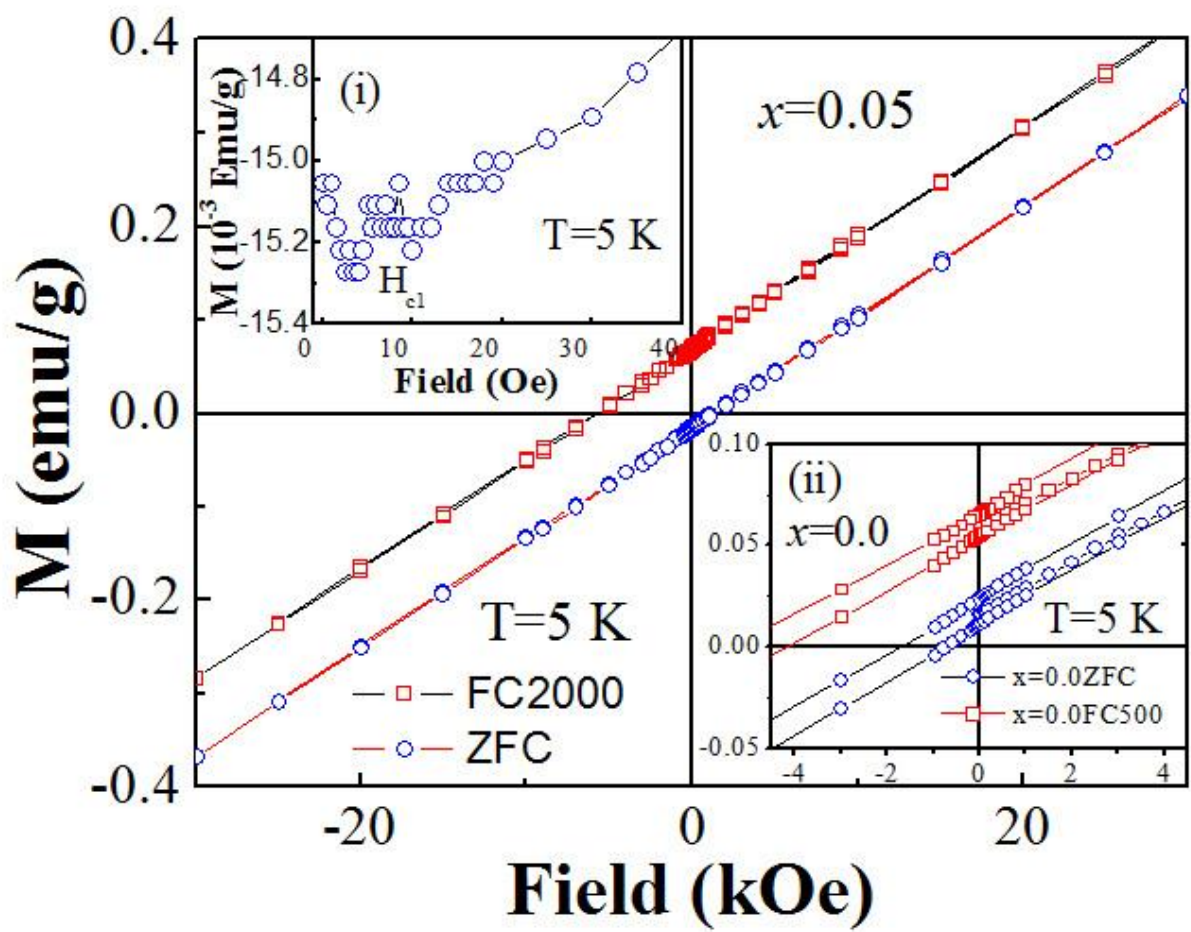

Figure 7

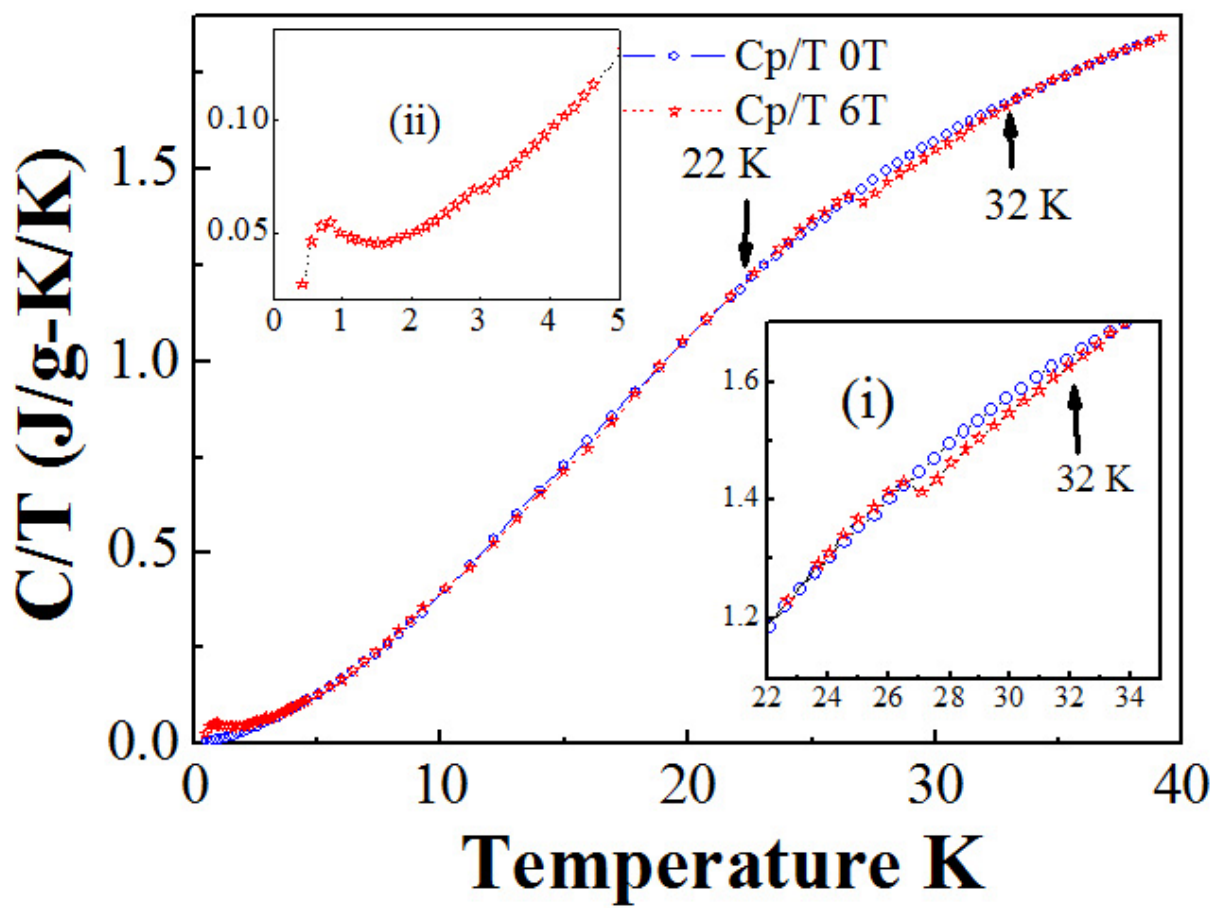

Figure 8

**Supporting data**

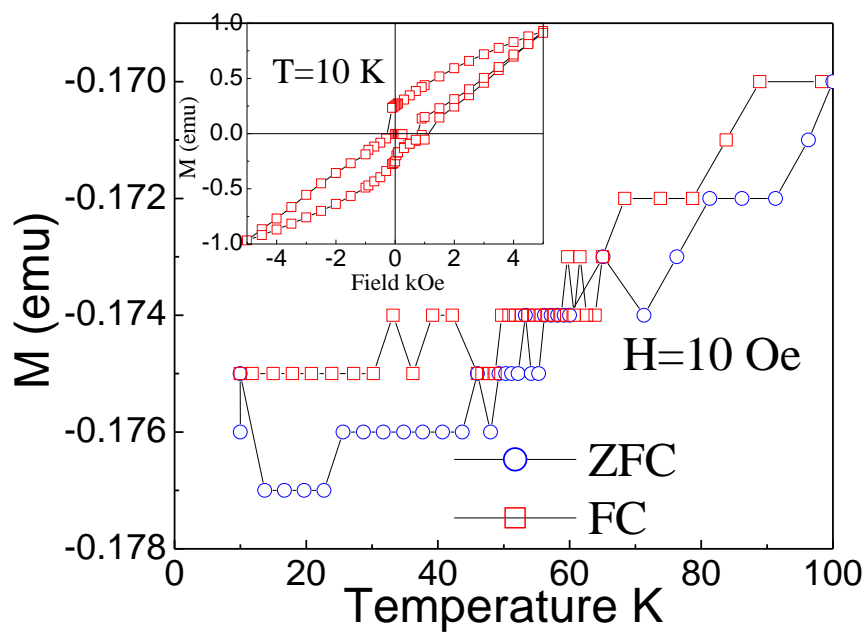

Fig. x1  The M-T and M-H curves of $SmBa_2Cu_3O_{6+\delta}$ single grain and ferromagnetic combination. The initial negative moment is observed to increase when ZFC field is applied as with the SrY2116 crystals. With a superconductor, it goes more negative.

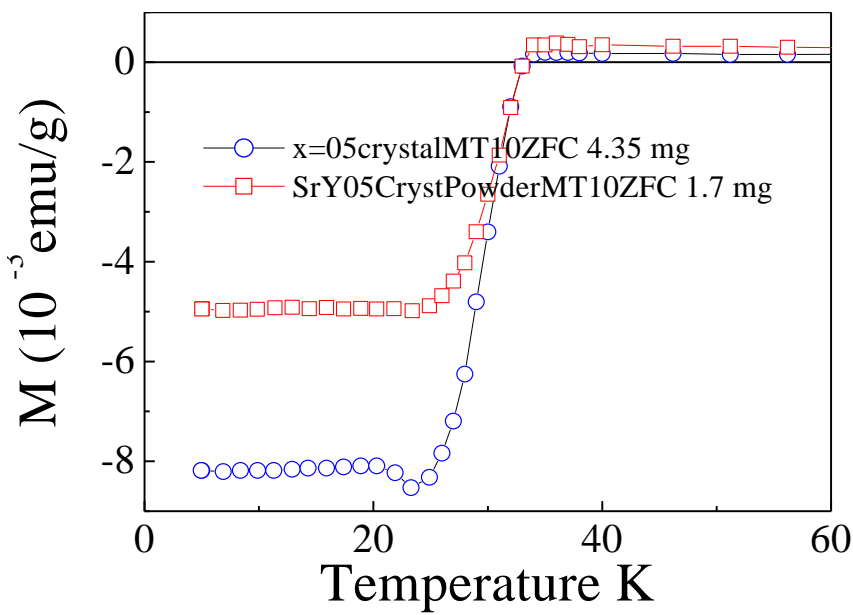

Fig. x2  M-T curves of Crystal and Powder. The magnitude of the powder is nearly 60% of the crystal.